\documentclass[twocolumn,trackchanges]{aastex701}
\usepackage{amsmath}
\usepackage{soul}
\usepackage{amssymb}
\usepackage{newunicodechar}
\newunicodechar{⁻}{\textsuperscript{-}}
\newunicodechar{−}{-}

\submitjournal{Publications of the Astronomical Society of the Pacific (PASP)}
\accepted{28 July 2026}

\begin{document}

\title{Statistical inference of fast radio burst environments using galaxy number density}

\author[orcid=0009-0001-9195-7494,sname='Vavillakula Venkataramana Rao']{Vignesh V.V. Rao$^{*}$}
\affiliation{Department of Physics, National Chung Hsing University, 145, Xingda Road, Taichung, 40227, Taiwan}
\email[show]{vigneshvavilla@gmail.com}  

\author[orcid=0000-0001-7228-1428,sname='Hashimoto']{Tetsuya Hashimoto$^{*}$} 
\affiliation{Department of Physics, National Chung Hsing University, 145, Xingda Road, Taichung, 40227, Taiwan}
\email[show]{tetsuya@phys.nchu.edu.tw}

\author[orcid=0000-0002-6821-8669]{Tomotsugu Goto}
\affiliation{Department of Physics, National Tsing Hua University, 101, Section 2. Kuang-Fu Road, Hsinchu, 30013, Taiwan}
\affiliation{Institute of Astronomy, National Tsing Hua University, 101, Section 2. Kuang-Fu Road, Hsinchu, 30013, Taiwan}
\email{itlhnis@gmail.com}

\author[orcid=0000-0002-1688-8708,sname='Yamasaki']{Shotaro Yamasaki}
\affiliation{Department of Physics, National Chung Hsing University, 145, Xingda Road, Taichung, 40227, Taiwan}
\email{shotarosyamasaki@gmail.com}

\author[orcid=0009-0005-9739-5540,sname='Madheshwaran']{Mohanraj Madheshwaran}
\affiliation{Department of Physics, National Chung Hsing University, 145, Xingda Road, Taichung, 40227, Taiwan}
\email{mohki96@hotmail.com}

\author[orcid=0000-0003-3747-9847,sname='Gajendran']{Sridhar Gajendran}
\affiliation{Institute of Astronomy, National Tsing Hua University, 101, Section 2. Kuang-Fu Road, Hsinchu, 30013, Taiwan} \affiliation{National Centre for Radio Astrophysics (NCRA–TIFR), Pune, 411007, India}
\email{}

\author[orcid=0000-0001-6010-714X,sname='Wada']{Tomoki Wada}
\affiliation{Frontier Research Institute for Interdisciplinary Sciences, Tohoku University, Sendai, Japan}
\affiliation{Astronomical Institute, Graduate School of Science, Tohoku University, Sendai, Japan}
\affiliation{Department of Physics, National Chung Hsing University, 145, Xingda Road, Taichung, 40227, Taiwan}\email{tomoki.wada@astr.tohoku.ac.jp}

\author[orcid=0000-0002-8560-3497,sname='Ho']{Simon C.-C. Ho}
\affiliation{Research School of Astronomy and Astrophysics, The Australian National University, Canberra, ACT 2611, Australia} \affiliation{Centre for Astrophysics and Supercomputing, Swinburne University of Technology, Hawthorn, VIC 3122, Australia} \affiliation{OzGrav: The Australian Research Council Centre of Excellence for Gravitational Wave Discovery, Hawthorn, VIC 3122, Australia} \affiliation{ASTRO3D: The Australian Research Council Centre of Excellence for All-sky Astrophysics in 3D, ACT 2611, Australia}
\email{}
\author[orcid=0009-0009-2940-0861,sname='Phan']{Terry Long Phan}
\affiliation{Institute of Astronomy, National Tsing Hua University, 101, Section 2. Kuang-Fu Road, Hsinchu, 30013, Taiwan}
\email{phanngoclong224@gmail.com}

\author[orcid=0000-0003-2792-4978,sname='Uno']{Yuri Uno}\affiliation{Department of Physics, National Chung Hsing University, 145, Xingda Road, Taichung, 40227, Taiwan}
\email{}

\author[orcid=0009-0004-9353-7065,sname='Chen']{Amos Y.-A.Chen}
\affiliation{Department of Physics, National Tsing Hua University, 101, Section 2. Kuang-Fu Road, Hsinchu, 30013, Taiwan} \email{}

\author[]{Hiroto Masaka}\affiliation{Mizusawa VLBI Observatory, University of Tokyo / National Astronomical Observatory of Japan (NAOJ)}
\email{}

\begin{abstract}

Fast radio bursts (FRBs) are bright, millisecond-duration radio transients of unknown origin. 
They are categorized as repeaters and non-repeaters, possibly indicating distinct progenitor types. 
However, validating this distinction is difficult because of the limited number of localized FRBs. 
Large-scale galactic environments can provide insight into the nature of the host galaxies of FRB and their progenitors. 
High-number-density regions are typically associated with old galaxies, whereas low-number-density regions are linked to young star-forming or less massive quiescent galaxies. 
In this study, we use galaxy number density to statistically assess the environments of 19
repeaters and 253 non-repeaters from CHIME Catalog 1, using galaxies from the WISE × PS1 catalog. 
A Kolmogorov–Smirnov (KS) test showed no significant difference between the two populations ($p_{KS} = 0.673$).
This result indicates that the statistical significance of the difference could depend on small-number statistics, highlighting the necessity of future FRB samples. Intriguingly, a comparison of FRBs with random galaxy fields suggests that FRBs may preferentially occur in underdense galactic environments with a median $p$-value ($p_{KS}$) = $2.84 \times 10^{-2}$ compared to random galaxy apertures. 


\end{abstract}


\keywords{\uat{Extragalactic astronomy}{506} ---  \uat{High Energy astrophysics}{739}--- \uat{Radio bursts}{1339}}

\section{Introduction}
\label{sec:Intro}

Fast radio bursts (FRBs) are bright ($\sim$ Jy), short-duration (millisecond), extragalactic radio transients \citep{Lorimer2007,Thornton2013}. 
Most FRBs happen at cosmological distances (galaxies outside our Milky Way). FRBs have a unique quantity called the dispersion measure (DM). DM is defined as the integral of the number density of free electrons along the line of sight (\ref{subsec:redshift} for details). 
Therefore, it can serve as a valuable cosmological probe for addressing key challenges in cosmology, including the missing baryon problem and the Hubble tension \citep[e.g.,][]{Petroff2019,Tzu-Yin2025,TC}.
However, their physical origin(s) remain uncertain, with a wide range of proposed scenarios \citep[e.g.,][]{Platts2019}, and thus understanding the origin of FRBs is one of the most important goals in astronomy. 

FRBs are classified as repeaters and non-repeaters based on the recurrence of their signals.
Observationally, an interesting distinction has emerged between these two classes in terms of their intrinsic temporal widths and spectral bandwidths, as found by CHIME \citep[e.g.,][]{Fonseca2020,CHIME2021cat1,Pleunis2021}.
These observations have raised the key question: do they have different progenitor types, with repeaters linked to longer-lived central engines and non-repeaters to cataclysmic events? \citep[e.g.,][]{Petroff2019,Cordes2019,Zhang2020,Tetsuya2020}.
Although some FRBs are localized very precisely within their host galaxies \citep[e.g.,][]{Marcote,Kirsten} and this provides a powerful way to understand their nature, the statistically conclusive evidence distinguishing different progenitor scenarios remains limited due to the small sample size of identified host galaxies \citep[e.g.,][]{Bhandari2022,Gordon2023}. 
CHIME plays a crucial role in understanding FRBs by providing a large number of FRB samples (around 530 sources){\footnote[1]{The CHIME/FRB Catalog 2 became available during the final stages of manuscript preparation and is therefore not included in the present analysis.}. 
However, the sample size of the host galaxies for this large dataset has been constrained by the localization capability. While CHIME/FRB Baseband observations \citep{CHIME_BASEBAND} combined with the KKO outrigger \citep{CHIME_KKO} achieve localization precisions of $\leq$10 arcsec and have enabled the construction of a new host-galaxy catalog, the resulting sample size remains substantially smaller than the full CHIME/FRB catalog \citep{CHIME2021cat1}. Given that the average redshift of FRBs in the CHIME sample is approximately $ z\sim$0.3, an accurate localization of $\sim$ 1 arcsec is required for host identification. 
Despite having nearly 100 localized FRB samples from the Australian Square Kilometer Array Pathfinder \citep[ASKAP;][]{Shannon2024ASKAP} Deep synoptic Array DSA, \citep[DSA;][]{Law2024DSA}, MeerKAT \citep{Jankowski2023MeerKAT}, and other telescopes, the sample size remains too small to draw any statistically significant conclusions. 
Therefore, to explore FRB progenitors effectively, an alternative approach that is independent of precise localization is needed.

%

The galaxy number density method circumvents the localization limitation mentioned above, as it probes large-scale structures ($\sim$ 6 Mpc scales). Also, the galaxy number density is well known to correlate with galaxy types: young late-type galaxy populations tend to reside in lower-density environments, while old early-type galaxy populations are in higher-density environments \citep[e.g.,][]{Dressler1980}. 
Therefore, the galaxy number density could allow us to statistically constrain the galaxy types of FRB hosts and their progenitor types.
Some previous works reported the galactic environments around FRBs. 
\citet{Connor2023} investigated the galactic environments surrounding two localized FRBs, FRB 20220914A and FRB 20220509G, to conclude that these FRBs occurred in highly dense environments.
However, previous work has focused mainly on high-density environments to investigate the DM contributions from the intersecting large-scale structures. For this purpose, host identification and spectroscopic redshift measurements are required, which have limited the number of samples in FRB statistical studies.
In this study, we propose a new statistical approach to constraining the host and progenitor types of FRBs using galactic environments around FRBs.
Our analysis is based on the CHIME Catalog 1 \citep{CHIME2021cat1}, which provides a uniform and statistically large FRB sample.
In contrast to the previous case studies, our method does not require spec-$z$ measurements of host galaxies of FRBs.
We use 253 non-repeaters and 19 
repeaters after selection criteria
(see \ref{sec:sample} for details). 
By using galaxy number density, we can investigate whether both repeater and non-repeater populations reside in the same galactic environment that could possibly link them to a common progenitor type. 
 
 A key advantage of our method is that it relies on statistical inference, which does not require accurate FRB localization of $\sim$ 1 arcsecond, but $\sim$ 10 arcminutes is sufficient because we calculate the number of galaxies inside the circle of fixed aperture (see
\ref{sec:sample} for details). Therefore, we can include unlocalized samples (see
\ref{sec:methodology} for details). 
Thus, a large enough sample of 19 repeaters and 253 non-repeaters is included in this work; hence, our analysis is less affected by the  of well-localized FRBs.

The paper is structured as follows. We describe the sample selection in \ref{sec:sample}. The methodologies involved in our analysis are discussed in \ref{sec:methodology}. The results and discussion are presented in \ref{sec:results} and \ref{sec:discussion}. We conclude in \ref{sec:conclusion}. Throughout this paper, a $\Lambda$CDM cosmology with parameters $H_0=70 
\: {\rm km\: s^{-1} \:Mpc^{-1}}$, $\Omega_{\rm m}=0.3$, and $\Omega_{\Lambda} = 0.7$ is used.
\section{Sample selections for FRBs and galaxies}
\label{sec:sample}

The non-repeater FRB population used in this study is sourced from CHIME Catalog 1 \citep{CHIME2021cat1}, while the repeater population is obtained from CHIME Catalog 1 \citep{CHIME2021cat1} and ``the golden sample" presented in \citet{CHIME2023repeaters}. 
CHIME Catalog 1 includes 472 one-off bursts and 62 bursts from 18 repeater sources. 
The WISE $\times$ PS1 photometric redshift (photo-$z$) catalog \citep{Beck2022} is used for the galaxy data sample. 
Its near-all-sky coverage and the availability of galaxy samples up to $z\sim0.8$ make this catalog ideal for our analysis. 
We select FRB and galaxy samples as detailed below (in \ref{subsec:FRB sample} and \ref{subsec:galaxy sample}).
Our selection criteria are based on those used by \citet{Tzu-Yin2025}, with specific modifications tailored to suit the objectives of this study. 
While \citet{Tzu-Yin2025} applied these criteria to compute densities of foreground galaxies for studying the IGM fluctuations, we adopt a redshift slice centered on each FRB's redshift (see \ref{sec:methodology} for details).
\subsection{FRB sample selection}
\label{subsec:FRB sample}
We applied the following selection criteria to select FRB samples in this work.
\begin{enumerate}
    \item FRB is located within the sky coverage of WISE $\times$ PS1 
    \item Galactic latitude $ \lvert \mathbf{b} \rvert $ $>$ 20$^\circ$
    \item Estimated redshift $z_{\rm FRB} < 0.8$ (see \ref{subsec:redshift} for details)
    \item FRB  samples with negative values of $z_{\rm FRB}$ were excluded.
    \item A visual inspection of the galaxy distribution surrounding each FRB was conducted to ensure data quality. 
\end{enumerate}
 The reason for following (2) is that the spatial distribution of the galaxy samples is significantly affected by Milky Way disk contamination and/or dust extinction. 
 (3) is to guarantee the completeness of the galaxy catalog (see \ref{subsec:galaxy sample} for details). 
 (4) selects extragalactic FRBs (see \ref{sec:methodology} for the details of the redshift calculation). 
As for (5), certain regions near the Galactic plane are masked in the WISE $\times$ PS1 catalog, leading to inaccurate density calculation in such regions. 
We removed FRB samples from our analysis when they were located in such masked regions.

 \subsection{Galaxy sample selection}
 \label{subsec:galaxy sample}
 The galaxies were selected with the following selection criteria:
 \begin{enumerate}
     \item Vega magnitude cut for W1 band of the WISE $\times$ PS1 samples (W1 $<$ 16.8 mag). %
     \item Galaxies were selected within a 100 $\times$ 100 Mpc$^{2}$ region, determined using the angular diameter distance at the redshift of each FRB sample.
     \item Galaxies were selected within a redshift slice which is created by using the redshift uncertainties of FRB (see \ref{subsec:redshift} for the details of redshift uncertainty calculation).
 \end{enumerate}
 (1) establishes a flux-limited sample, ensuring the completeness of the sample given that observational completeness declines beyond 16.8 mag \footnote[1]{\url{https://wise2.ipac.caltech.edu/docs/release/allsky/expsup/sec6_3a.html}}. 
 (2) is to collect sufficient reference galaxies. 
 A 100 $\times$ 100 Mpc$^{2}$ region is adopted as galaxies in this physical scale are unlikely to be associated with FRB's environments, but they represent randomly selected galaxy number densities.

\section{Methodology}
\label{sec:methodology}
The methodology used in this study includes estimating the FRB redshift using the Macquart relation \citep{Macquart2020} (see \ref{subsec:redshift} for more details). Once the redshift is determined, we calculate the uncertainties in the redshift value and use it as a redshift slice around each FRB (see \ref{subsec:slice} for more details). Later, we calculate the galaxy number density for each repeater and non-repeater FRB and use these values to understand the galactic environment around those FRBs (see \ref{subsec:density} for more details). 
\subsection{Redshift estimation of FRB samples}
\label{subsec:redshift}

The redshift of each FRB source is measured using the dispersion measure (DM). 
DM is defined as the integral of the number density of free electrons along the line of sight. 
The observed dispersion measure (DM$_{\rm{obs}}$) is contributed by different components:
\begin{equation}
\label{eq:DM_FRB}
    {\rm DM}_{\rm obs} ={\rm DM}_{\rm MW}+ {\rm DM}_{\rm IGM}(z)+{\rm DM}_{\rm host}/(1+z),
\end{equation}
 which include the DM due to the host galaxy of the FRB (DM$_{\rm{host}}$), DM contribution from Milky Way galaxy (DM$_{\rm{MW}}$), and DM due to the intergalactic medium (DM$_{\rm{IGM}})$ \citep{Ioka2003,Inoue2004,Zhou2014,Macquart2020}. 
 The DM$_{\rm{MW}}$ further comprises the DM due to warm Galactic halo (DM$_{\rm{MW,halo}})$, \citep{Prochaska2019, YT2020} and cold Galactic interstellar medium (DM$_{\rm{MW,ISM}}$). 
 The contribution from DM$_{\rm{MW,ISM}}$ is computed using ``fruitbat.FRB" \citep{Batten2019Fruitbat}, a Python package that applies the YMW16 Galactic DM model \citep{YMW16}.
The DM$_{\rm IGM}$ is estimated by assuming the contributions of  DM$_{\rm{MW,halo}} = 65 $ pc cm$^{-3}$ \citep{Prochaska2019} and DM$_{\rm host} = 50$ pc cm$^{-3}$ \citep{Shannon2018} for simplicity. 
As DM$_{\rm IGM}$ depends on redshift ($z$) of FRB sources \citep{Zhou2014}, we numerically solve Equation (\ref{eq:DM_FRB}) to obtain the redshift of each source, denoted $z_{\rm FRB}$. 
Figure \ref{fig:redshift} shows the redshift distribution of FRB sources from CHIME Catalog 1 \citep{CHIME2021cat1} before applying the selection criteria described in Section \ref{subsec:FRB sample}. 
A red solid vertical line shows the redshift cut at $z_{\rm FRB}=0.8$. 
The majority of the FRB samples are well within the redshift cut.
In this work, we use the formalization of the Macquart relation \citep{Macquart2020} by \citep{Zhou2014}. The actual Macquart relation \citep{Macquart2020} may differ slightly from this, depending on the telescopes \citep{Shin2022}, potentially introducing a bias in our analysis. However, the potential bias does not significantly affect our results because the same Macquart relation is applied to both repeating and non-repeating FRBs.

\subsection{Redshift slice for galaxy samples}
\label{subsec:slice}
After estimating $z_{\rm FRB}$, we determine its uncertainty in order to define an appropriate redshift slice for galaxy selection (see \ref{subsec:galaxy sample}).

The major contribution to the uncertainty of the DM-derived redshift comes from the line-of-sight fluctuation of DM$_{\rm IGM}$.
Therefore, we use the DM$_{\rm IGM}$ fluctuation to estimate the redshift uncertainty. 
The dispersion in DM$_{\rm IGM}$ is calculated for each source using the following equation given by  \citet{Cordes2022},    see also \citep{Takahashi2021},
\begin{equation}
\label{eq:sigma_IGM}
    \sigma_{{\rm DM}_{\rm IGM}}(z) = [{\rm DM}_c {\rm DM}_{\rm IGM}(z) ]^{1/2},
\end{equation}
where $\sigma_{{\rm DM}_{\rm IGM}}(z)$ represents the root-mean-square (rms) scatter in the dispersion measure contribution from the intergalactic medium (IGM) as a function of redshift $z$, due to cosmic variance. ${\rm DM}_c =50$ pc cm$^{-3}$ is a constant value for approximation, and ${\rm DM}_{\rm IGM}$ is the contribution of the dispersion measure from the intergalactic medium. 
Incorporating the line-of-sight fluctuations of DM$_{\rm IGM}$ given by equation \eqref{eq:sigma_IGM}, we replace ${\rm DM}_{\rm IGM}(z)$ with ${\rm DM}_{\rm IGM}(z) \pm \sigma_{{\rm DM}_{\rm IGM}}(z)$ in equation \eqref{eq:DM_FRB}, and solve for $z$ to estimate the uncertainty of the redshift. 
The redshift uncertainty associated with each FRB was utilized to define an appropriate redshift slice for galaxy selection. 
The galaxies lying outside this slice are excluded from the analysis (see also \ref{subsec:density}).

We note that the DM-derived redshift is influenced by the line-of-sight fluctuations in DM$_{\rm IGM}$. 
Therefore, the DM-derived redshift would be underestimated (overestimated) if an FRB is located behind less-dense (denser) galaxy-number-density environments than the cosmic average. 
Underestimating or overestimating the redshift could affect density calculation. 
To mitigate this effect, we adopt the redshift slice based on the upper and lower bounds of the line-of-sight fluctuations in DM$_{\rm IGM}$ as mentioned above.
This redshift slice ensures that the true redshift of an FRB is included within the slice.

The median photometric redshift error of our galaxy sample is ($\Delta z_{photo} = 3.03 \times 10^{-2}$), which is significantly smaller than the median redshift error of FRBs, $(\Delta z_{FRB} = 2.81 \times 10^{-1})$, estimated from DM. To reliably include galaxies associated with FRB environments, we adopt redshift slices based on FRB redshift uncertainties, making the contribution from galaxy photometric redshift errors negligible by comparison. Moreover, near the high-redshift edge of the redshift slices, photometric redshift uncertainties may sometimes cause background galaxies to be included in the redshift slice, or galaxies within the slice to be shifted into the background. The same happens for the low-redshift edge of the slice. However, if the galaxy densities in the foreground and background are similar, these shifts have little impact on the estimated galaxy number density. As a result, the measured density remains largely insensitive to the photometric redshift errors  \citep[e.g.,][]{Tzu-Yin2025}.

\begin{figure}
    \centering
    \includegraphics[width=1.0\columnwidth]{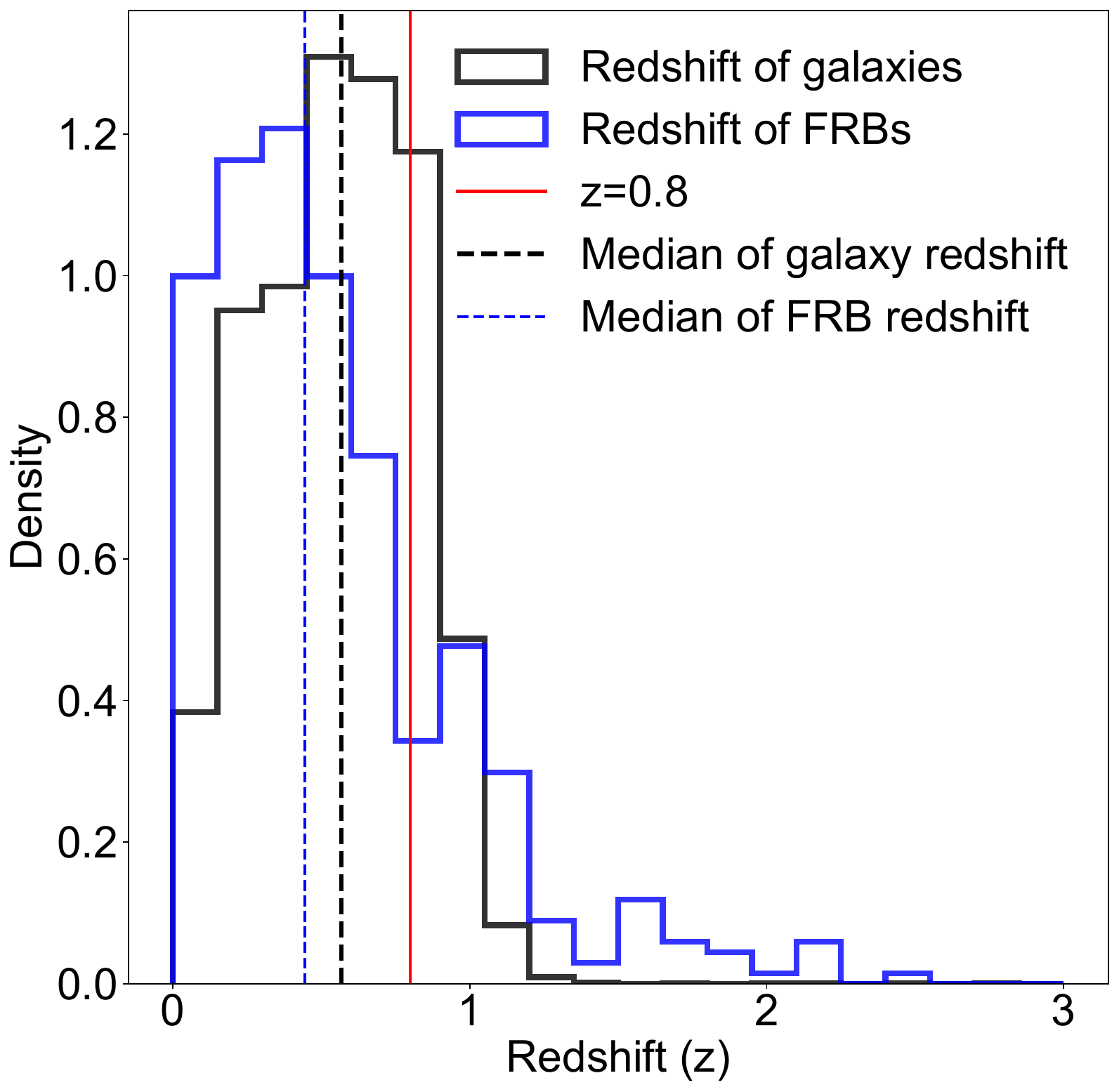}
    \caption{Estimated redshift distribution for the CHIME Catalog 1  and the redshift distribution of 10,000 randomly selected galaxies from the PS1 catalog. 
    The mean redshift of FRBs is 0.54 with a standard deviation of ($ \pm 1\sigma $) $=$0.43. 
    The red vertical line denotes the redshift cut that we follow in our analysis. 
    The black and blue solid lines correspond to the redshifts of the galaxies and FRBs, respectively. 
    The black and blue vertical dashed lines correspond to their median values, respectively.}
    \label{fig:redshift}
\end{figure}
 
\subsection{Calculation of galaxy-number density surrounding FRBs}
\label{subsec:density}

\begin{figure}[htp!]
    \centering
    \includegraphics[width=1.0\columnwidth]{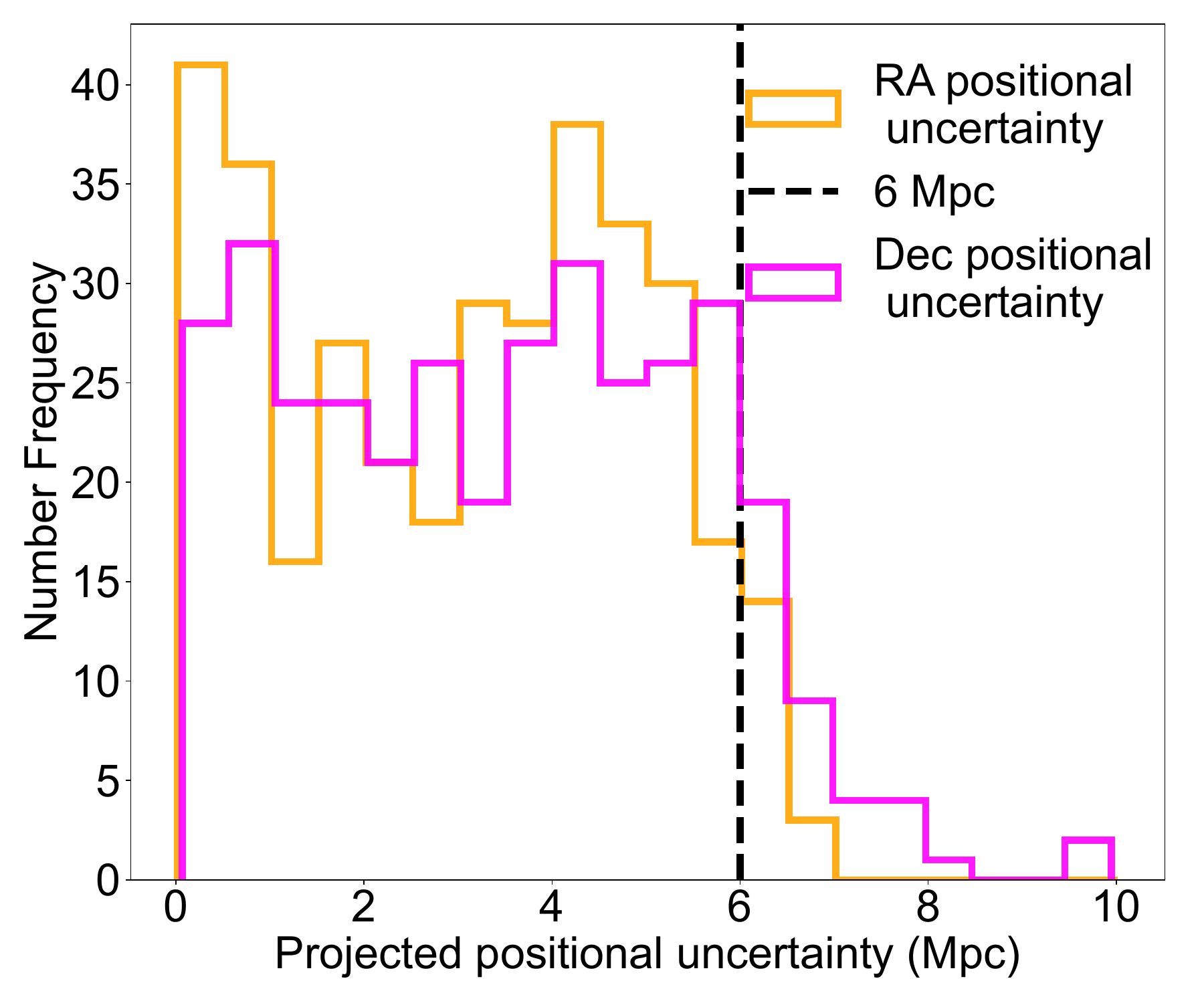}
    \caption{ Histogram of positional uncertainties of all the CHIME FRBs converted to projected physical distances (Mpc). The black dashed-vertical line corresponds to the 6 Mpc aperture radius adopted in calculating the galaxy number densities (see \ref{sec:methodology} for details)}.
    \label{fig:positional}
\end{figure}

Figure \ref{fig:positional} shows a histogram of FRB's positional errors converted to projected physical distances along Right Ascension (RA) and Declination (Dec). We found that approximately 89\% of our samples exhibit positional errors within 6 Mpc.
Consequently, we adopt a 6 Mpc aperture radius to calculate galaxy number densities in this analysis.

For the galaxy samples, we constructed a 100 $\times$ 100 Mpc$^2$ square region centered on each FRB sample.
First, we converted the Galactic latitude and Galactic longitude in the field of each FRB sample into the projected comoving distance using the FRB's redshift. 
Then, we constructed a 100 $\times$ 100 Mpc$^2$ region around them and used the coordinates of the region's endpoints to select the galaxy samples. 
The galaxy samples inside this region are filtered using a Vega magnitude cut for the W1 band of WISE $\times$ PS1 (W1 $<$ 16.8) magnitude (Section \ref{subsec:galaxy sample}).
Then, the redshift-slice cut (Section \ref{subsec:galaxy sample}) is applied based on the error values that we obtain during our redshift estimation (Section \ref{subsec:redshift}). 
This step excludes foreground galaxies that are distant and not associated with the FRB redshift, ensuring they do not bias the environmental density calculations.

Figure \ref{fig:redshiftcut} is an example of our samples, showing a spatial galaxy distribution inside a 100 $\times$ 100 Mpc$^{2}$ region after the redshift slice cut. 
Once the redshift slice is constructed, we estimate the galaxy number density by creating fixed-radius apertures. 
We count the number of galaxies within the aperture and divide it by the aperture area to derive the galaxy number density. 
Since the positional uncertainties of most FRB samples are well inside the 6 Mpc (see Figure \ref{fig:positional}), we define the aperture radius to be 6 Mpc, thereby including the positional uncertainties of the sources. 
\begin{figure}[h!]
    \centering
    \includegraphics[width=1.0\columnwidth]{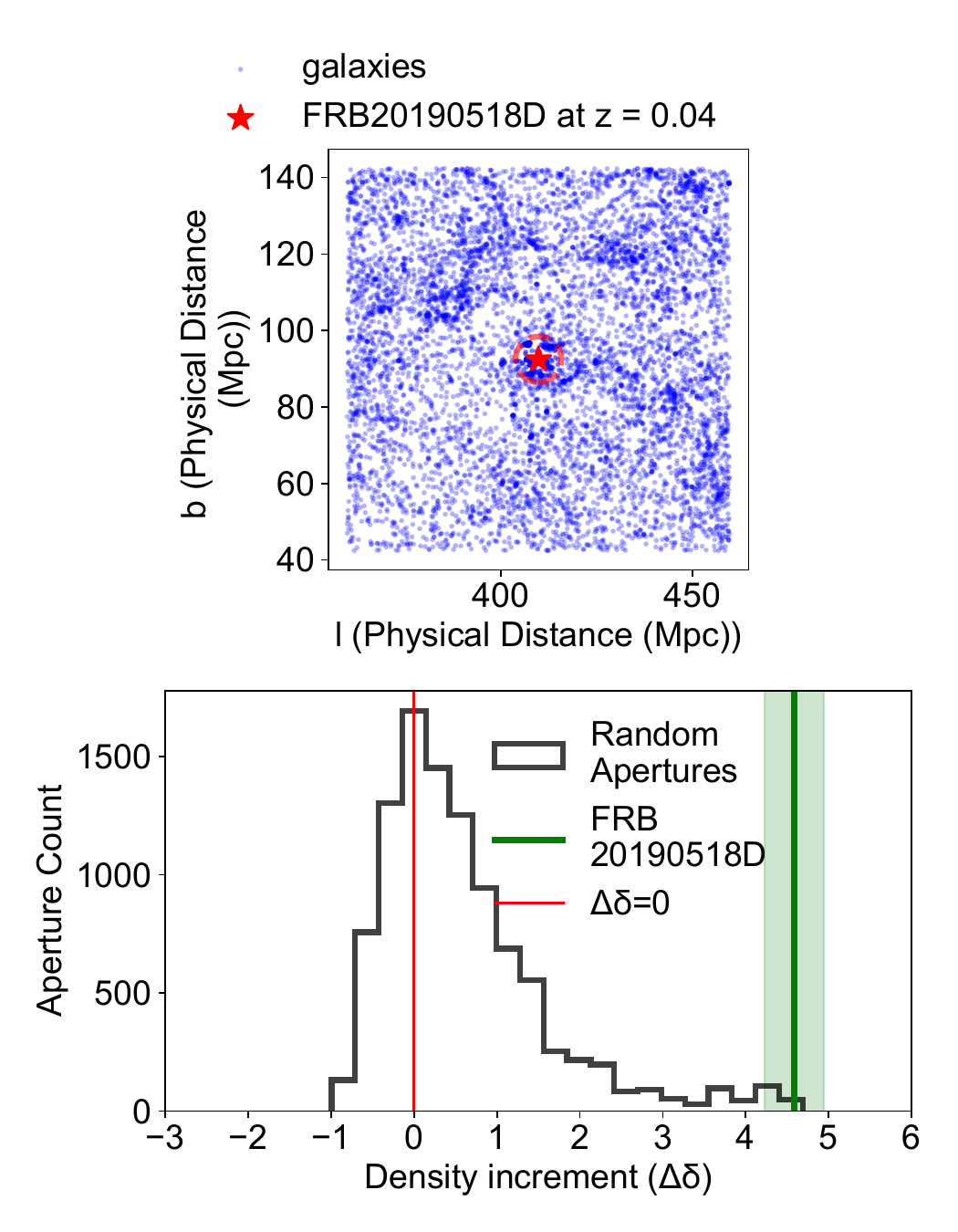}
    \caption{(Top) An example of the galaxy distribution around FRB 20190518D. The X and Y axes correspond to the Galactic coordinates ($l$, $b$) in physical distances. The entire size of the plot corresponds to 100 $\times$ 100 Mpc$^{2}$ around the FRB. The red circle indicates the 6 Mpc aperture radius. The red star is the location of FRB 20190518D. (Bottom) A histogram representing the density increment of the random
apertures in the 100 $\times$ 100 Mpc$^{2}$ field around the FRB 20190518D. The green line
corresponds to the normalized density increment around the location of the FRB. 
The value of the normalized number density around FRB 20190518D is 4.6 times higher than that of the reference number densities. The green-shaded region corresponds to the Poisson error of the galaxy number count.}
\label{fig:redshiftcut}
\end{figure}

The absolute value of the projected galaxy number density highly depends on the depth of the survey and the FRB's redshift \citep[e.g.,][]{Tzu-Yin2025}, as the fainter galaxies are difficult to detect in shallower surveys or at higher redshifts.  
To eliminate this effect, we calculate reference projected-number densities for each FRB sample as follows. 
We construct $N$ random apertures, each with the same radius as the one used for the FRB, within a 100 $\times$ 100 Mpc$^{2}$ region centered on each FRB source. 
For an aperture radius of 6 Mpc within a 100 $\times$ 100 Mpc$^{2}$ FRB field, we adopt $N=10^4$ random apertures to ensure statistically robust estimates.
The reference number densities are calculated by counting the number of galaxies within each of the $N$ random apertures in each FRB field.
Then, the reference densities are normalized by their standard deviation.
Let $\rho_i$ be the reference density value for the $i$-th aperture ($i=1,\cdots,N$), and $\sigma_{\rho}$ denotes the standard deviation of the density distribution:

\begin{equation}
    \delta_{i} = \frac{\rho_i}{\sigma_{\rho}},
\end{equation}
where $\delta_{i}$ represents the normalized galaxy number density relative to the standard deviation of the densities in the random apertures. 
Normalization of number densities helps to mitigate Malmquist bias in the sample. 
The same normalization is applied to each FRB sample, which provides us with a clear comparison of the FRB number density to the reference number densities.
After normalization, we construct a histogram of the reference densities for each FRB sample to identify the peak density. 
To quantify the relative offset from the typical field density, we shift the normalized reference densities such that they are measured relative to the peak value. 
We refer to these as density increments, denoted by ${\Delta}\delta_{i}$:
\begin{equation}
\label{eq:sigma}
    \Delta\delta_{i} = \delta_{i}-\delta_{\rm random,\,peak},
\end{equation}
where $\delta_{\rm random,\, peak}$ represents the peak of the normalized number density distribution for random apertures.
We select the peak of the random density distribution as the origin because it directly represents the deviation of the FRB density environment from the random peak in sigma units. By subtracting from the peak, we can see how much higher/lower the density is compared to the majority of random apertures.
The same normalization and subtraction are performed for the 
density of each FRB sample ($\Delta\delta_{\rm FRB}$) as follows.
\begin{equation}
\label{eq:sigma_FRB}
    \Delta\delta_{\rm FRB} = \delta_{\rm FRB}-\delta_{\rm random,\,peak},
\end{equation}
\noindent where $\delta_{\rm FRB}$ denotes the normalized number density of a single aperture centered on each FRB position.
Accordingly, FRBs with $\Delta \delta_{\rm FRB} > 0$ are classified as an overdense environment while those with $\Delta \delta_{\rm FRB} < 0$ are considered to be an underdense environment. Figure \ref{fig:redshiftcut} is an example of our samples, in an overdense ($\Delta \delta_{\rm FRB}>0$) galactic environment, showing the density increment calculated from the normalized reference densities according to Equation \eqref{eq:sigma_FRB}. We repeat this procedure for all of the FRB samples and individually measure their $\Delta\delta_{\rm FRB}$.
Afterwards, we use the notation, $\Delta\delta$, to represent the density increment values for either FRB or randomly selected samples (see also Section \ref{subsec:comparision of FRB and rand}).


\begin{figure}
    \centering
    \includegraphics[width=1.0\columnwidth]{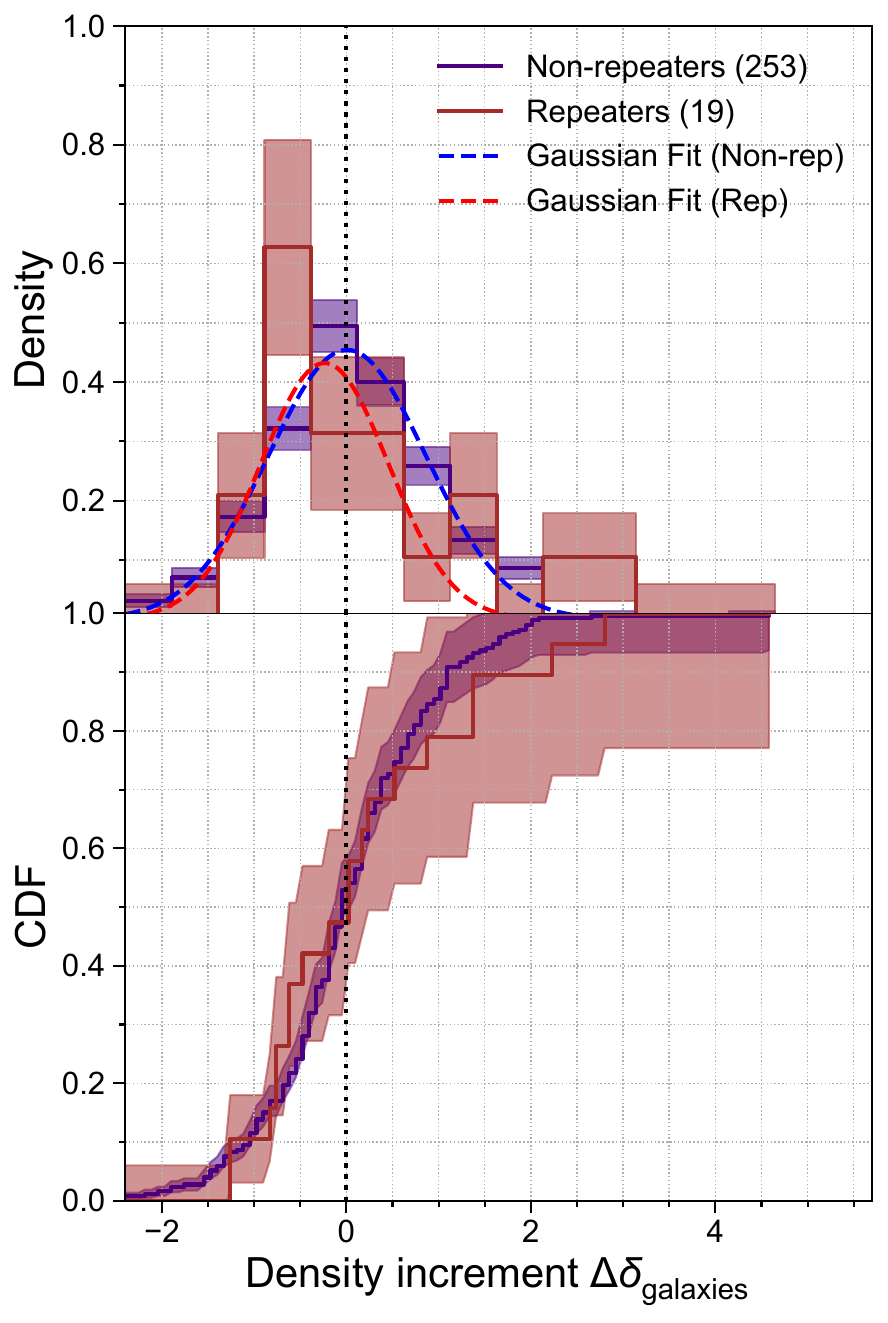}  
    \caption{Histograms (Top) and cumulative distribution functions (bottom) of the normalized density increment values for both repeaters and non-repeaters. The indigo line corresponds to the normalized density increment of non-repeaters, and the brown line corresponds to the normalized density increment of repeaters. The blue dashed line corresponds to the Gaussian fit for non-repeaters, and the red dashed line corresponds to the Gaussian fit for repeaters.}
    \label{fig:density}
\end{figure}

\begin{figure}
    \centering
    \includegraphics[width=1.0\columnwidth]{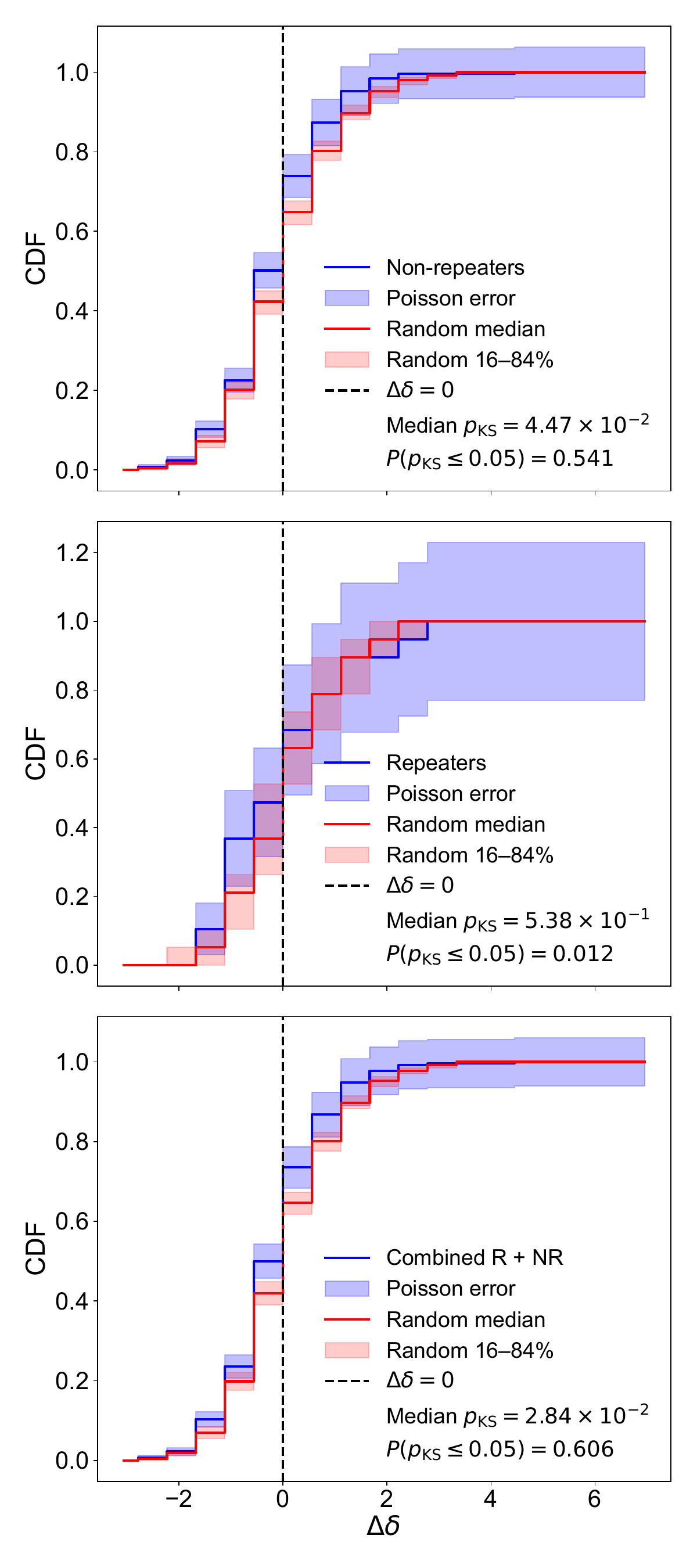}
    \caption{Cumulative distribution functions of $\Delta\delta$ for observed FRBs and Monte
Carlo random-aperture realizations. The blue curves show the observed FRB
distributions, and the blue shaded regions indicate Poisson counting
uncertainties. The red curves show the bin-wise median CDFs of the Monte Carlo
random-aperture realizations, while the red-shaded regions indicate the
16th--84th percentile ranges across the Monte Carlo trials. The vertical dashed
line marks $\Delta\delta=0$.}
    \label{fig:nr_and_rand}
\end{figure}

\section{Results}
\label{sec:results}

In this section, we present the key findings of our analysis. Our results can be summarized into three main findings, each discussed in a separate subsection. Section~\ref{subsec:comparison} examines the galactic environments of repeating and non-repeating FRBs. 
Finally, Section \ref{subsec:comparision of FRB and rand} compares the galaxy number densities around FRBs with those derived from randomly placed apertures. This comparison is designed to determine whether the observed galaxy number densities reflect a physical association or are consistent with random selection effects. Throughout this work, we do not privilege any particular outcome and treat all scenarios with equal weight.

\subsection{Comparison between repeaters and non-repeaters}
\label{subsec:comparison}
Figure \ref{fig:density} shows histograms (top) and cumulative distribution functions (CDFs) (bottom) of $\Delta\delta_{\rm FRB}$ to compare the galactic environments of repeaters and non-repeaters.
A Kolmogorov-Smirnov (KS) test was conducted to assess whether both samples originate from the same distribution. 
The resulting $p$-value for the KS test is $p_{KS} = 0.673$.

The $p$-value indicates statistically no significant difference in the galactic environments between the two populations. To examine whether the histogram peaks are consistent within their errors, we applied a Gaussian fit to each histogram. We compared the histogram peaks with the associated error regions. 
We found that the fitted peak positions are $\mu = -0.24 \pm 0.15$ for repeaters.
In the case of non-repeaters, the fitted peaks are  $\mu = 0.0088 \pm 0.050$. $\mu$ represents the peak of the histogram of each population. 

Therefore, we do not see a significant difference between the histogram peaks of repeaters and non-repeaters. One possible reason for the absence of a significant difference is the small sample size, as discussed in Section \ref{subsec:difference between rep and non-rep}.

\subsection{Comparison of galaxy number densities between randomly selected apertures and FRB }
\label{subsec:comparision of FRB and rand}

In this section, we test whether the FRB galaxy number density distribution is consistent with random aperture selection. For the non-repeater population, each Monte Carlo realization is constructed by selecting the galaxy number density around one random aperture out of $10^4$ random aperture values for each FRB. This results in a single random density sample of the same size as the observed 253 non-repeaters sample. We perform a KS test between the observed FRB's density distribution and the one set of random realizations to compute a $p$-value. These procedures are repeated to generate 1000 sets of Monte Carlo realizations and $p$-values. This approach avoids treating the repeated random-aperture samples as a single fully independent aggregated catalog. The top panel of Figure \ref{fig:nr_and_rand} shows the cumulative distribution function (CDF) of $\Delta\delta$, where $\Delta\delta$ represents the density increment of non-repeaters and random apertures. The blue solid line shows the observed non-repeater distribution, while the blue shaded region indicates the Poisson uncertainty. The red solid line represents the bin-wise median CDF of the Monte Carlo random-aperture realizations, and the red shaded region indicates the 16th–84th percentile range across the Monte Carlo trials. For the non-repeater sample, the row-wise Monte Carlo KS analysis yields a median KS $p$-value of $p_{KS} = 4.47 \times 10^{−2}$, with 54.1\% of Monte Carlo trials satisfying $p_{KS} \leq 0.05$. This result indicates a statistically significant deviation of the non-repeater density distribution from the random-aperture expectation. Compared with the random-aperture CDF, the observed non-repeater CDF is shifted leftward, indicating that a larger fraction of non-repeaters occupy lower $\Delta\delta$ values. This trend suggests that non-repeaters preferentially reside in relatively underdense environments compared with randomly selected regions. The middle panel of Figure \ref{fig:nr_and_rand} shows the same comparison for the repeater sample. Here, $\Delta\delta$ represents the density increment of repeaters. For the repeater sample, the median KS $p$-value is $p_{KS} = 5.38 \times 10^{−1}$, and only 1.2\% of the Monte Carlo trials satisfy $p_{KS} \leq 0.05$. Therefore, the repeater density distribution shows no statistically significant difference from the random-aperture distribution. However, this result may be strongly limited by the small-number-statistics of the repeater sample and therefore may not fully represent the true nature of repeater FRB environments. The bottom panel of Figure \ref{fig:nr_and_rand} shows the comparison for the combined FRB sample, including both repeaters and non-repeaters. Here, $\Delta\delta$ represents the density increment of the combined FRB sample. For this combined sample, the row-wise Monte Carlo KS analysis yields a median KS $p$-value of $p_{KS} = 2.84 \times 10^{−2}$, with 60.6\% of Monte Carlo trials satisfying $p_{KS} \leq 0.05$. The combined FRB CDF is also shifted leftward relative to the median random-aperture CDF, indicating that the full FRB sample contains a larger fraction of sources at lower $\Delta\delta$ values than expected from random-aperture selection. This result indicates a statistically significant deviation of the combined FRB population from the random-aperture expectation, mainly driven by the non-repeater population.

If repeater and non-repeater samples originate from different redshift distributions, the environmental comparisons may be affected by redshift-dependent selection effects. To examine whether our final results are affected by redshift-dependent selection effects, we performed a redshift-matching analysis between the repeater and non-repeater samples. Specifically, we calculated the fraction of repeaters in each redshift bin and selected the same fraction of non-repeaters within the corresponding bin. This procedure ensured that the redshift distributions of the repeater and non-repeater populations were matched. The resulting matched non-repeater sample contains 149 non-repeaters. Using this matched sample, we recalculated the density increment,  $\Delta\delta$, and repeated the full statistical analysis. The results show no statistically significant difference between repeaters and non-repeaters, with a $p$-value of $p_{KS} = 5.7 \times 10^{-1}$.  In contrast, the comparison between non-repeaters and the random distribution yields a statistically significant difference, with a $p$-value of $p_{KS} = 1.6 \times 10^{-2}$. Similarly, when the full FRB sample (repeaters and non-repeaters combined) is compared with random apertures, the difference remains statistically significant, with a $p$-value of $1.5 \times 10^{-2}$. These results demonstrate that our main conclusions remain unchanged even after accounting for possible redshift-dependent selection effects through redshift matching. Therefore, the observed environmental trends are unlikely to be driven by differences in the redshift distributions of repeaters and non-repeaters.

\section{Discussion}
\label{sec:discussion}
In this section, we interpret the results of our analysis. Section \ref{subsec:difference between rep and non-rep} presents the interpretation of the results in the case of comparison between repeaters and non-repeaters.
Finally, Section \ref{ssubsec:preference} examines the results obtained from comparisons between FRBs and random galaxy apertures. 
\subsection{Comparison between repeater and non-repeater populations}
\label{subsec:difference between rep and non-rep}

The $p$-value of $p_{KS}=0.673$
shown in \ref{subsec:comparison} indicates that there is no statistically significant difference between repeaters and non-repeaters.
One possible reason for this result is the small-number-statistics contributed by the repeater sample. To see solid differences or similarities in the repeater and non-repeater distributions, increasing the number of samples using the upcoming telescopes, including the CHIME Catalog 2 \citep{CHIMECAT2}, Bustling Universe Radio Survey Telescope in Taiwan (BURSTT) \citep{Lin2023burstt,Ling2023}, is important. Such future telescopes could help us place stronger, more conclusive constraints on the galactic environments around both repeater and non-repeater populations.
On the other hand, it is also plausible that both repeaters and non-repeaters originate from similar galactic environments; that is, the two populations may not be intrinsically distinct but instead represent manifestations of the same underlying population \citep[e.g.,][]{Ravi2019,Yamasaki2024}.

\subsection{Preferential occurrence of FRB in underdense regions compared to random galaxy positions}
\label{ssubsec:preference}
There is a well-established empirical relationship between the galaxy number density and the galaxy type \citep[e.g.,][]{Dressler1980}.
Galaxies in high-density environments may contain old galaxy populations, such as elliptical and S0 galaxies, whereas underdense regions contain star-forming galaxies \citep[e.g.,][]{Dressler1980} and less massive quiescent galaxies.}
 
From the top panel of Figure \ref{fig:nr_and_rand}, we can see that the galaxy number density of non-repeaters is shifted towards the underdense region compared to that of randomly selected apertures. 
The observed preference for underdense environments may suggest an association with young star-forming host galaxies or lower-mass quiescent galaxies rather than with old, massive galaxies \citep{Dressler1980,Cucciati2017}, although this interpretation remains statistical in nature and does not constitute a direct measurement of host-galaxy properties.

In the case of galaxy number densities between repeaters and random selection distribution, there is no significant difference between the two populations in the middle panel of Figure \ref{fig:nr_and_rand}. The median $p$-value is $p_{KS}= 5.38 \times 10^{−1}$. 
The main reason for this $p$-value could be a very small repeater sample, which could potentially affect the $p$-value.

When we combine both the repeater and the non-repeater distributions and compare them with the galaxy number density values of a randomly selected distribution (bottom panel of Figure \ref{fig:nr_and_rand}), the median $p$-value ($p_{KS}) = 2.84\times10^{-2}$ using the K-S test.
We find that the overall distribution of FRBs is significantly shifted toward underdense environments relative to random apertures. Based on the well-established correlations among galaxy environment, stellar mass, and morphology \citep{Dressler1980,Cucciati2017}, this trend suggests that FRBs are statistically more likely to be associated with star-forming (late-type) galaxies or relatively low-mass quiescent galaxies, rather than with massive quiescent galaxies. We emphasize that our interpretation is statistical in nature and is not intended to serve as a proxy for direct host-galaxy searches.
Also, our results do not preclude the possibility of FRB occurrence in massive quiescent hosts; rather, such environments appear to be comparatively less favored than lower-density, field-like environments.


\subsection{Sensitivity to the Assumed Host DM Contribution}
In this work, following \citep{Macquart2020},  we adopt a fiducial host galaxy dispersion measure of DM$_{\rm host} = 50 $ pc cm$^{-3}$, motivated by theoretical considerations presented in  \citep{Macquart2020}. To evaluate the robustness of our results to this assumption, we repeat the analysis using an alternative value of DM$_{\rm host} = 100 $ pc cm$^{-3}$. We find that the comparison between non-repeater environments and random apertures becomes marginally consistent with the null hypothesis ($p_{KS} = 5.18 \times 10^{-2}$), whereas the combined FRB sample continues to show a statistically significant difference relative to random positions ($p_{KS} = 1.5 \times 10^{-2}$).  Therefore, the conclusion on the comparison between the non-repeater sample and random apertures could be subject to systematic uncertainties associated with the assumed DM$_{\rm host}$. Therefore, it is important to better constrain DM$_{\rm host}$ through independent observational measurements. For instance, scattering-based analyses \citep[e.g.,][]{Cordes2019,ocker2022,TC} and H $\alpha$-based estimates \citep[e.g.,][]{Bernales-Cortes2025,Tendulkar2017} provide complementary approaches to empirically determine the host galaxy contribution to the dispersion measure.

\subsection{Robustness Test with Alternative Background Scales}

We adopt 100 Mpc $\times$ 100 Mpc because the typical large-scale structure could be $\sim$ 100 Mpc scale. Background regions smaller than 100 Mpc $\times$ 100 Mpc could be affected by the cosmic variance. For instance, a smaller region could include only high-density large-scale structures. In such a case, the correct density increment cannot be estimated. To avoid this effect, we use 100 Mpc $\times$ 100 Mpc as a fiducial parameter set. To assess the sensitivity of our results to the choice of background region, we performed a test using a 150 Mpc $\times$ 150 Mpc region, motivated by the physical scale of the baryon acoustic oscillations \citep{Eisenstein2005}. In this case, statistically significant differences exist between FRBs and the random distribution (non-repeaters vs random $p_{KS} =  7.36 \times 10^{-4}$; repeaters vs random $p_{KS} = 2.07 \times 10^{-1}$; FRBs (combined) vs random $p_{KS} = 2.60 \times 10^{-4}$).

\subsection{Sensitivity of the Results to Aperture Size}

To test the robustness of our results, we explored alternative aperture sizes in addition to our fiducial model. For a smaller aperture radius of 2 Mpc, the available sample size is substantially reduced, leaving only 38 non-repeater FRBs in the analysis. Despite the reduced sample size, the KS test against random apertures yields a statistically significant difference between the non-repeater FRB population and the random aperture distribution ($p_{KS} = 2.31 \times 10^{-4}$). Similarly, when the repeater and non-repeater samples are combined into a single FRB population, we again detect a statistically significant difference relative to the random apertures ($p_{KS} = 6.93 \times 10^{-4}$). However, consistent with our fiducial analysis, the repeater population alone does not exhibit a statistically significant difference from the random distribution. We further examined an intermediate aperture radius of 4 Mpc. In this case, we also find a statistically significant difference between the non-repeater FRB population and the random apertures ($p_{KS} = 7.45 \times 10^{-4}$), while the repeater population again remains statistically consistent with the random distribution. When the repeater and non-repeater samples are combined, the full FRB population continues to show a statistically significant difference relative to the random apertures ($p_{KS} = 8.36 \times 10^{-4}$). Overall, the conclusions obtained from both the 2 Mpc and 4 Mpc aperture analyses remain fully consistent with those of our fiducial model, indicating that our main results are robust against reasonable variations in aperture size.

\subsection{Broad FRB redshift uncertainties and projection effects}

The large FRB redshift uncertainties inferred from DM can potentially dilute environmental contrasts by introducing projection effects along the line of sight. To quantitatively assess this effect, we performed an additional analysis using 10 FRB samples residing in overdense regions higher than $\sim2$ $\Delta\delta$. For these samples, the photo-z distributions of galaxies within 6 Mpc from FRB locations indicate clear peaks with much narrower distributions than the DM-derived redshift slices. This suggests that they are associated with overdense structures. Supposing that these FRB samples are physically associated with the overdense structures, we defined a new redshift slice based on the median photo-z $ \pm 1\sigma$ of galaxies within 6 Mpc for each FRB sample. Therefore, the new redshift slices are narrower than the original DM-derived redshift slices.

We recalculated the normalized galaxy number density increments ($\Delta\delta$) using these new redshift slices and compared the results with our original estimates based on the broader DM-derived redshift slices. We found that the effect of the large redshift slice can both dilute and enhance the density contrast. In some cases, foreground/background field galaxies are included in the density calculation, diluting the density contrast. In other cases, foreground/background overdense regions along the line of sight can be included within the large redshift slice, enhancing the density contrast. Based on the above-mentioned analysis, we found that this uncertainty is about 0.7 $\Delta\delta$. Therefore, the absolute values of our $\Delta\delta$ measurements are subject to this uncertainty. However, our major focus is not the absolute value but the relative difference between FRB densities and those in the random apertures. The same redshift slice is used for both the FRB and random-aperture samples, indicating that FRBs may preferentially occur in lower-density environments compared with random apertures. Therefore, this conclusion would not be significantly affected by the uncertainty of 0.7 $\Delta\delta$.

\section{Conclusion}
\label{sec:conclusion}
The primary objective of this study is to understand galactic environments without precise localization of FRB sources. 
An important motivation for examining the distinction or similarities between the galactic environments is to constrain the progenitor types for FRBs. Although previous studies have constrained the host galaxy types of FRBs to infer progenitor types, their sample sizes are limited. Using the CHIME and Pan-STARRS data, we improved the sample size by a factor of two (253 non-repeaters and 19 repeaters), compared to previous works \citep{Connor2023}, which shows the significance of our research.  

In the case of analyzing galaxy number densities 
our result (see section \ref{sec:results} for details) (see Figure \ref{fig:density} for more details) showed no significant difference between the galactic environments of repeaters and non-repeaters. 

From the results, it is evident
that the statistical difference between the galactic environments of repeaters and non-repeaters depends on the number of samples.
This can be well constrained by increasing the number of samples from future FRB observations.

When comparing the galactic environments of FRBs with random galaxy apertures, we find a statistically significant difference with a KS test $p$-value $=$ $ 2.84\times10^{-2}$.
Our results indicate that FRBs preferentially occur in low-density galactic environments (see the bottom panel of Figure \ref{fig:nr_and_rand}). This trend suggests that FRBs are statistically more likely to be associated with star-forming (late-type) galaxies or relatively low-mass quiescent galaxies, rather than with massive quiescent galaxies. Nevertheless, we note that this interpretation is statistical in nature and does not preclude the possibility of FRB occurrence in massive quiescent hosts. Instead, such environments appear to be comparatively less favored than lower-density, field-like environments. Also, we emphasize that our interpretation is statistical in nature and is not intended to serve as a proxy for direct host-galaxy searches.

This result provides an important observational constraint on the nature of FRB host galaxies and their progenitor models. We note some caveats regarding potential detection biases of FRBs in dense galactic environments. In particular, we note that the statistical significance of the observed environmental trend exhibits some dependence on the assumed host-galaxy dispersion measure, indicating sensitivity to systematic uncertainties in the redshift and dispersion measure decomposition. Therefore, independent constraints on the host-galaxy dispersion measure, for example through scattering-based or H$\alpha$-based estimations, are important for improving the robustness of future analyses. Also, increased scattering and dispersion smearing due to a larger DM can lead to pulse broadening, potentially reducing detection efficiency. While the present work represents a best-effort analysis with the currently available data, a detailed quantitative assessment of these effects is beyond the scope of this study. Despite these limitations and the currently limited number of confirmed repeaters, our work provides a foundation for future studies using CHIME Catalog 2 and upcoming facilities.

\section*{Acknowledgement}
We express our deepest gratitude to Murthadza Aznam from the Department of Physics, Faculty of Science, Universiti of Malaya, Kuala Lumpur, Malaysia for his valuable insights on using the Golden Catalog to improve the sample size of repeaters. We extend our gratitude to Tzu-yin Hsu from Swinburne University of Technology for her support and insights on working with the PS1 STRM database. We are grateful to the anonymous referee for their careful review of the manuscript and for their insightful comments and constructive suggestions, which have helped improve the clarity and quality of this work. TG acknowledges the support of the National Science and Technology Council of Taiwan (NSTC) through grants 113-2112-M-007 -006, 113 -2927-I-007 -501, 113-2123-M-001 -008. 
TH acknowledges the support of NSTC through grants 113-2112-M-005-009-MY3, 114-2123-M-001-002-, and 111-2112-M-005-018-MY3.
SY acknowledges the support from NSTC through grant numbers 113-2112-M-005-007-MY3 and 113-2811-M-005-006-. T.W. acknowledges the Grants-in-Aid for Scientific Research No. 25KJ0024, 25K17378 from the Ministry of Education, Culture, Sports, Science and Technology (MEXT) of Japan. We acknowledge use of the CHIME/FRB Public Database, available at \url{https://www.chime-frb.ca/} from the CHIME/FRB Collaboration \citep{CHIME2021cat1}. This research made use of Astropy, a community-developed core Python package for Astronomy \citep{2018AJ....156..123A}. 

\section*{Data Availability}
All data analyzed in the current study are available from the
CHIME/FRB Public Database (https://www.chime-frb.ca/) and
CHIME/FRB Open Data (https://chime-frb-open-data.github.io/).
Custom codes and the files to reproduce galaxy density analysises will be
made available upon reasonable request to the corresponding author.

\bibliography{PASPsample701}{}
\bibliographystyle{aasjournalv7}



\end{document}